\documentclass[twocolumn,showpacs,nofootinbib,amsmath,amssymb,prd]{revtex4}

\usepackage{hyperref}
\usepackage{mathrsfs}
\usepackage{latexsym}
\usepackage{amsmath}
\usepackage{amsbsy}
\usepackage{amssymb}
\usepackage{epsfig}
\usepackage{graphicx}
\usepackage{dcolumn}
\usepackage{graphicx}
\usepackage{bm}
\usepackage{natbib}

\newcommand{\da}{\partial}
\newcommand{\al}{\alpha}
\newcommand{\be}{\beta}
\newcommand{\de}{\delta}

\newcommand{\ep}{\varepsilon}
\newcommand{\si}{\sigma}

\newcommand{\om}{\omega}

\newcommand{\Om}{\Omega}

\newcommand{\M}{\mathcal{M}}

\newcommand{\U}{\mathcal{U}}

\newcommand{\Lie}{\mathcal{L}}

\newcommand{\Q}{\mathcal{Q}}

\newcommand{\J}{\mathcal{J}}

\newcommand{\E}{\mathcal{E}}
\newcommand{\B}{\mathcal{B}}
\newcommand{\eem}{EM}
\newcommand{\fluid}{F}
\newcommand{\Tem}{T^{\eem}}
\newcommand{\rhoem}{\rho^{\eem}}
\newcommand{\Pem}{P^{\eem}}
\newcommand{\jem}{j^{\eem}}
\newcommand{\Piem}{\Pi^{\eem}}
\newcommand{\Tmat}{T^{\fluid}}
\newcommand{\rhomat}{\rho^{\fluid}}
\newcommand{\Pmat}{P^{\fluid}}
\newcommand{\jmat}{j^{\fluid}}
\newcommand{\Pimat}{\Pi^{\fluid}}

\newcommand{\real}{\mathbb{R}}
\newcommand{\rAH}{r_{\pm}}

\begin{document}

\title{Gravitational collapse of spherically symmetric plasmas in Einstein-Maxwell spacetimes}

\author{P. D. Lasky}
	\email{paul.lasky@sci.monash.edu.au}, 

\author{A. W. C. Lun}
	\email{anthony.lun@sci.monash.edu.au}

\affiliation{Centre for Stellar and Planetary Astrophysics\\
		School of Mathematical Sciences, Monash University\\
		Wellington Rd, Melbourne 3800, Australia}
\received{\today}
	
		\begin{abstract}
			We utilize a recent formulation of a spherically symmetric spacetime endowed with a general decomposition of the energy momentum tensor [{\it Phys. Rev. 
D} {\bf 75} 024031 (2007)] to derive equations governing spherically symmetric distributions of electromagnetic matter.  We show the system reduces to the Reissner-Nordstrom spacetime in general, spherically symmetric coordinates in the vacuum limit.  Furthermore, we show reduction to the charged Vaidya spacetime in non-null coordinates when certain equations of states are chosen.  A model of gravitational collapse is discussed whereby a charged fluid resides within a boundary of finite radial extent on the initial hypersurface, and is allowed to radiate charged particles.  Our formalism allows for the discussion of all regions in this model without the need for complicated matching schemes at the interfaces between successive regions.  As further examples we consider the collapse of a thin shell of charged matter onto a Reissner-Nordstrom black hole.  Finally, we reduce the entire system of equations to the static case such that we have the equations for hydrostatic equilibrium of a charged fluid.  
		\end{abstract}
		
		\maketitle
		
\section{INTRODUCTION}
Recently \citet{lasky07} reduced Einstein's field equations governing a spherically symmetric distribution of a completely general decomposition of the energy-momentum tensor to a system conducive to an initial value problem.  It was shown that under certain equations of state, the system reduced to the Vaidya spacetime in non-null coordinates.  Furthermore, when all energy-momentum terms vanished the system reduced to the Schwarzschild spacetime in general coordinates.  This formulation enables all regions of the spacetime to be written in a single coordinate system with equations of state tending to certain values at the interfaces between the regions.  Therefore, the use of junction conditions at the interfaces is not required as one simply establishes an initial hypersurface and lets the system evolve via the governing set of equations.  

The present paper highlights the generality of the fluid elucidated in \cite{lasky07} by incorporating electromagnetic (EM) fields into the spacetime.  In this way we are doing plasma physics in full general relativity and we show that the equations have familiar Newtonian analogues.  The role of EM fields in fully relativistic gravitational collapse has been the topic of much interest over recent years \cite{ori90a,barreto99,fayos03}.  Of particular interest is the formation of naked singularities \cite{stein-schabes85,ori91,krasinski06}, the formation and evolution of charged neutron stars within the context of full general relativity \cite{ray03,ghezzi05,ghezzi07}, and the collapse of charged shells of matter onto existing black holes \cite{cherubini02}.  We utilize a formalism which puts general, spherically symmetric charged fluids into an initial value setting on a spacelike hypersurface such that all of these issues can be discussed either analytically or numerically.  This paper concentrates on the formulation of the plasma equations, with an emphasis on the reduction of the full set of equations to various solutions of charged fluids in spherical symmetry, including the investigation of the collapse of infinitely thin shells.  

Known charged fluid solutions in general relativity include the charged Vaidya and Reissner-Nordstrom (RN) spacetimes.  The generalized Birkhoff's theorem implies the RN spacetime is the unique, spherically symmetric vacuum solution endowed with an electric field \cite{misner73}.  The charged Vaidya solution \cite{lindquist65,israel67,plebanski68,bonnor70} is a generalization of the Vaidya metric which describes null radiation.  Both the Vaidya and charged Vaidya spacetimes are usually exhibited by using null coordinates, and indeed the mathematical form of the metric in these coordinates is rather simple \cite{hussain96,wang99}.  However, the physical interpretation of the fluid and the matching of the fluid to other spacetimes is fraught with difficulties \cite{bonnor89,herrera04}.  In \cite{lasky07}, we re-cast the Vaidya spacetime in non-null coordinates such that it is conducive to an initial value formulation and showed that this made the matching of spacetimes significantly simpler.  In this article we extend the result to include the charged Vaidya spacetime in spacelike coordinates and further discuss the matching of this to an interior charged fluid and also to an exterior RN solution.    

Another example we elucidate in this paper is that of an infinitely thin charged shell falling onto an RN black hole.  We are able to describe this entire scenario in terms of a single line element and hence as a single solution of the Einstein field equations.  Furthermore, the differential equation governing the collapsing charged shell is reduced to quadrature, implying the dynamics of the entire spacetime is known.  

The Einstein-Maxwell equations describe the coupling of a gravitational field to an EM field.  These comprise the Einstein field equations whereby the energy momentum tensor has a contribution from the Faraday tensor.  The resultant energy momentum tensor can be represented as simply the sum of two fields \cite{misner73}
\begin{align}
T_{\mu\nu}=\Tmat_{\mu\nu}+\Tem_{\mu\nu}.\label{Totalsum}
\end{align}
Throughout the article a superscript ``$F$'' represents entities relating to the fluid contribution to the energy momentum tensor, $\Tmat_{\mu\nu}$, and a superscript ``$EM$'' relates quantities to the EM contribution to the energy momentum tensor, $\Tem_{\mu\nu}$.  In section \ref{sec1} of this article we review the form of $\Tem_{\mu\nu}$ and the Maxwell equations in a covariant formulation.  Furthermore, in this section we decompose Maxwell's equations and $\Tem_{\mu\nu}$ into spatial and temporal components by exploiting the $3+1$ formalism for general relativity.  Section \ref{sec2} then reviews the necessary components of the general decomposition of $T_{\mu\nu}$ and the equations derived in \cite{lasky07}.   We point out the existence of the generalization of the Lorentz force law when EM is coupled to the Einstein field equations.  This equation is often neglected when studying Einstein-Maxwell fields, however it is a necessary component of the system of equations when modelling physically realistic scenario's.  Section \ref{sec3} then puts together the results of the two preceding sections to derive the reduced field equations governing a spherically symmetric distribution of charged matter.  In section \ref{RN} we show how the system reduces to the RN spacetime in the appropriate limit and section \ref{Vad} shows the derivation of the relevant equations of state and the reduction of the system to the charged Vaidya spacetime. 
We move to a slightly different application in section \ref{shell} to discuss in detail a shell of charged matter collapsing onto a black hole.  We show the entire system is given by a single line element, where the differential equation of the shell reduces to quadrature.  Finally, in section \ref{static} we reduce the system to the static sub-case and hence derive the equations of hydrostatic equilibrium for a charged fluid ball.  

Throughout the article we use a $+2$ signature and coordinates are denoted by $x^{\mu}=(t,r,\theta,\phi)$ unless otherwise stated.  Geometrized units are used, whereby $c=G=1$.  Greek indices range from $0\ldots3$, Latin from $1\ldots3$ and all other conventions follow \cite{misner73}.

\section{Covariant Electromagnetism}\label{sec1}
We provide here a brief introduction to the relevant components of the $3+1$ formalism for general relativity.  Begin by defining a vector field, $n^{\mu}$, which is timelike and normalized, $n^{\al}n_{\al}=-1$.  Further demanding this satisfies Frobenius theorem,
\begin{align}
n_{[\mu}\nabla_{\nu}n_{\si]}=0,
\end{align}
where square brackets denote antisymmetrization and $\nabla_{\mu}$ is the unique four-metric connection, implies the normal vector is hypersurface forming.  Further define a projection tensor, $\perp_{\mu\nu}$, which projects all quantities onto the spacelike hypersurfaces formed by the normal vector,
\begin{align}
\perp_{\mu\nu}:=g_{\mu\nu}+n_{\mu}n_{\nu},
\end{align}   
where $g_{\mu\nu}$ is the four-metric.

The energy momentum tensor for EM fields is expressed in terms of the anti-symmetric Faraday tensor \cite{landau75}
\begin{align}
\Tem_{\mu\nu}=\frac{1}{4\pi}\left(F_{\al\mu}{F^{\al}}_{\nu}-\frac{1}{4}g_{\mu\nu}F_{\al\be}F^{\al\be}\right).\label{Tem}
\end{align}
An observer moving in the direction of the normal vector will then measure the electric and magnetic field intensities respectively as
\begin{align}
E_{\mu}:=F_{\mu\al}n^{\al}\qquad\textrm{and}\qquad B_{\mu}:=\frac{1}{2}\ep_{\mu\al\be}F^{\al\be},
\end{align}
where $\ep_{\mu\nu\si}$ is the three-Levi Civita pseudo-tensor.  One can show from the anti-symmetric nature of the Faraday tensor and the spatial nature of the three-Levi Civita pseudo tensor that $E^{\al}n_{\al}=B^{\al}n_{\al}=0$, implying the electric and magnetic fields are spatial.  The Faraday tensor can be uniquely decomposed onto and orthogonal to the spacelike hypersurfaces specified by its normal vector.  After some algebra we can show
\begin{align}
F_{\mu\nu}=\ep_{\mu\nu\al}B^{\al}-2E_{[\mu}n_{\nu]}.\label{FFFFun}
\end{align}
Using the above information, equation (\ref{Tem}) can be irreducibly decomposed resulting in the energy momentum tensor being expressed in terms of the electric and magnetic 
fields
\begin{align}
\Tem_{\mu\nu}=\rhoem n_{\mu}n_{\nu}+\Pem\perp_{\mu\nu}+2\jem_{(\mu}n_{\nu)}+\Piem_{\mu\nu},\label{Tem2}
\end{align}
where
\begin{align}
\rhoem:&=\frac{1}{8\pi}\left(E_{\al}E^{\al}+B_{\al}B^{\al}\right),\label{rhoem}\\
\Pem:&=\frac{1}{24\pi}\left(E_{\al}E^{\al}+B_{\al}B^{\al}\right),\label{Pem}\\
\jem_{\mu}:&=\frac{1}{4\pi}\ep_{\mu\al\be}E^{\al}B^{\be},\label{Poynting}\\
\Piem_{\mu\nu}:&=\frac{1}{4\pi}\left[\frac{1}{3}\perp_{\mu\nu}\left(E_{\al}E^{\al}+B_{\al}B^{\al}\right)-E_{\mu}E_{\nu}-B_{\mu}B_{\nu}\right].\label{Piem}
\end{align}
We note the vector $\jem_{\mu}$ is the EM Poynting vector \cite{tsagas05} which is a measure of the energy flow in the spacetime.    

Maxwell's equations are expressed covariantly in terms of the Faraday tensor according to \cite{landau75}
\begin{align}
\nabla_{[\si}F_{\mu\nu]}=&0,\label{Max1}\\
\nabla^{\al}F_{\mu\al}=&4\pi\J_{\mu}.\label{Max2}
\end{align}
Here, $\J_{\mu}$ is the charge-current four-vector, which can be decomposed according to
\begin{align}
\J_{\mu}=i_{\mu}+\ep n_{\mu},\label{Jdef}
\end{align}
where $i_{\mu}:={\perp_{\mu}}^{\al}\J_{\al}$ is the three-current density and $\ep:=-n^{\al}\J_{\al}$ is the charge density.  By taking the divergence of both sides of equation (\ref
{Max2}), we obtain the conservation law for the charge-current four-vector, $\nabla^{\al}\J_{\al}=0$.  Decomposing this with respect to the normal implies
\begin{align}
\Lie_{n}\ep-K\ep+\left(D^{\al}+\dot{n}^{\al}\right)i_{\al}=0,\label{Lieep}
\end{align}   
where $\Lie_{n}$ is the Lie derivative operator with respect to the normal vector field, $K$ is the trace of the extrinsic curvature, $D_{i}$ is the unique three-metric connection 
associated with $\perp_{ij}$ and $\dot{n}_{\mu}:=n^{\al}\nabla_{\al}n_{\mu}$ is the observers four-acceleration.  Equation (\ref{Lieep}) is an identity if Maxwell's equations are satisfied.  However, it can be used if one is given either an arbitrary charge distribution or the three-current density to determine the other unknown quantity.  

Decomposing equation (\ref{Max1}) with respect to the normal vector provides two equations
\begin{align}
D^{\al}B_{\al}=&0,\label{divB}\\
\Lie_{n}B_{\mu}-KB_{\mu}+2{K_{\mu}}^{\al}B_{\al}=&-\ep_{\mu\al\be}\left(D^{\al}+\dot{n}^{\al}\right)E^{\be},\label{LieB}
\end{align}
where $K_{\mu\nu}$ is the extrinsic curvature.  Decomposing equation (\ref{Max2}) gives
\begin{align}
D^{\al}E_{\al}=&4\pi\ep,\label{divE}\\
\Lie_{n}E_{\mu}-KE_{\mu}+2{K_{\mu}}^{\al}E_{\al}=&\ep_{\mu\al\be}\left(D^{\al}+\dot{n}^{\al}\right)B^{\be}-4\pi i_{\mu}.\label{LieE}
\end{align}
Equations (\ref{divB}-\ref{LieE}) are the more familiar version of Maxwell's equations in curved spacetime.  

To complete the description of EM in general relativity we now derive the Lorentz force law.  We expand the divergence of the EM component of the energy momentum tensor, $\nabla^{\al}\Tem_{\mu\al}$, with respect to the definition in terms of the Faraday tensor (\ref{Tem}).  Then by using both the anti-symmetric nature of the Faraday tensor and the four-dimensional Maxwell equations, (\ref{Max1}) and (\ref{Max2}), we can show
\begin{align}
\nabla^{\al}\Tem_{\mu\al}=-F_{\mu\al}\J^{\al}.\label{Lorentz1}
\end{align}
Conservation of energy momentum in general relativity is a statement that the divergence of the total energy momentum tensor vanishes.  By expanding this in terms of the EM and fluid components, equation (\ref{Totalsum}), and substituting equation (\ref{Lorentz1}) we find the conservation of energy momentum in general relativity gives
\begin{align}
\nabla^{\al}\Tmat_{\mu\al}=F_{\mu\al}\J^{\al}.\label{Lorentz2}
\end{align} 
This is a four-dimensional equation which can be decomposed into three constraint equations on the hypersurface and a single propagation equation.  This decomposition involves taking the ${\perp_{\mu}}^{\al}\nabla^{\be}\Tmat_{\al\be}$ component and the  $n^{\al}\nabla^{\be}\Tmat_{\al\be}$ component respectively.  However, the left hand sides of these equations requires the explicit knowledge of the decomposition of the fluid component of the energy momentum equation which shall be discussed in the next section.  

Summarily, equations (\ref{Lieep}-\ref{LieE}) provide the full set of Maxwell's equations governing EM fields in curved spacetime.  These couple back to gravitation through the conservation of energy momentum tensor given by equation (\ref{Lorentz2}).

\section{fluid formulation}\label{sec2}
\subsection{Covariant equations}
In \cite{lasky07} we used a generic decomposition of the total energy momentum tensor given by
\begin{align}
T_{\mu\nu}=&\rho n_{\mu}n_{\nu}+P\perp_{\mu\nu}+2j_{(\mu}n_{\nu)}+\Pi_{\mu\nu},\label{Ttot}
\end{align}
where $\rho$ and $P$ are the total energy density and total isotropic pressure respectively, $j^{\mu}$ is interpreted loosely as a heat flux vector and $\Pi_{\mu\nu}$ as an anisotropic stress tensor. 

We use the same decomposition for the fluid component of the energy momentum, $\Tmat_{\mu\nu}$, as we used for the total energy momentum tensor, i.e. equation (\ref{Ttot}).  
However, we denote each quantity decomposed from the fluid contribution to the total energy momentum tensor with a superscript ``$\fluid$''.  The total energy momentum components are now given in terms of their constituent fluid and EM contributions by using equations (\ref{Totalsum}), (\ref{rhoem}-\ref{Piem}) and (\ref{Ttot}), and we 
find 
\begin{align}
\rho&=\rhomat+\frac{1}{8\pi}\left(E_{\al}E^{\al}+B_{\al}B^{\al}\right),\label{rhotot}\\
P&=\Pmat+\frac{1}{24\pi}\left(E_{\al}E^{\al}+B_{\al}B^{\al}\right),\label{Ptot}\\
j_{\mu}&=\jmat_{\mu}+\frac{1}{4\pi}\ep_{\mu\al\be}E^{\al}B^{\be},\label{jtot}\\
\Pi_{\mu\nu}&=\Pimat_{\mu\nu}+\Piem_{\mu\nu},\label{Pitot}
\end{align}
where $\Piem_{\mu\nu}$ is given by equation (\ref{Piem}).

Before moving on to the spherically symmetric reduction of the equations, we discuss the conservation of energy momentum equations which lead to a generalized Lorentz force law.  As mentioned above, we decompose equation (\ref{Lorentz1}) onto and orthogonal to the spacelike hypersurface formed by the normal vector.  The first equation can be found from the ${\perp_{\mu}}^{\al}\nabla^{\be}\Tmat_{\al\be}$ part of the equation, which after some algebra and using equations (\ref{FFFFun}), (\ref{Jdef}) and (\ref{Ttot}) can be shown to give
\begin{align}
\Lie_{n}\jmat_{\mu}-&K\jmat_{\mu}+\left(D^{\al}+\dot{n}^{\al}\right)\Pimat_{\mu\al}+D_{\mu}\Pmat\nonumber\\
&+\left(\rhomat+\Pmat\right)\dot{n}_{\mu}=\ep E_{\mu}+\ep_{\mu\al\be}i^{\al}B^{\be}.\label{Lorentz}
\end{align}
This is a covariant representation of the Lorentz force law coupled to the gravitational field, which can be seen intuitively by considering an example of the flow of some current.  Here, $i_{\mu}$ describes the flow of the charge and $\jmat_{\mu}$ describes the flow of the ``fluid particles''.  As the charge is tied to the fluid, these two quantities essentially describe the same phenomena.  Therefore, the term $\Lie_{n}\jmat_{\mu}$ describes a timelike derivative of a flow of fluid particles, i.e. the rate of change of momentum of fluid particles.  Furthermore, the right hand side is the electric field plus the wedge product of the three-current density and the magnetic field.  \citet{feynman64} states that this force law is required in classical EM theory {\it because having all the electric and magnetic fields doesn't tell us anything until we know what they do to the charges} (see chapter 18-3 of \cite{feynman64}).     

The second term from the decomposition of the conservation of energy momentum is given by $n^{\al}\nabla^{\be}\Tmat_{\al\be}$.  Again after some manipulation this becomes
\begin{align}
\Lie_{n}\rhomat-\left(\rhomat+\Pmat\right)K+&\left(D^{\al}+\dot{n}^{\al}\right)\jmat_{\al}\nonumber\\
&-\Pimat_{\al\be}K^{\al\be}=-E_{\al}i^{\al}.\label{energy}
\end{align}
The left hand side of this equation is equivalent to the continuity equation of fluid dynamics.  The term on the right hand side describes the {\it electrical energy for the plasma} \cite{priest82}.  This equation has a Newtonian analogue in plasma physics as an alternative form of the heat equation and it represents the fact that the gain in energy of the material is due to many different facets including heat flux, electrical energy and the work done by the pressure (see equation 2.40 of \citet{priest82}).
 
\subsection{Spherical Symmetry}
Spherical symmetry implies the density and pressure are just functions of the temporal and radial coordinates.  We can also write the heat flux vector as
\begin{align}
j_{\mu}=\left[\be j(t,r),\,j(t,r),\,0,\,0\right].\label{jform}
\end{align}
Furthermore, we define a spatial tensor according to
\begin{align}
{P_{i}}^{j}:=\textrm{diag}\left[-2,\,1,\,1\right].
\end{align} 
Now, the anisotropic stress tensor, $\Pi_{\mu\nu}$ is symmetric, spatial ($\Pi_{\mu\al}n^{\al}=0$) and trace-free (${\Pi_{\al}}^{\al}=0$).  It is trivial to show with spherical symmetry we can always write
\begin{align}
\Pi_{\mu\nu}:&=\Pi(t,r)P_{\mu\nu},\label{Piform}
\end{align}
where $\Pi(t,r)$ is the distinct eigenvalue of $\Pi_{ij}$.

In \cite{lasky07} we further began with a spherically symmetric line element expressed in $3+1$ form
\begin{align}
dS^{2}=-\al^{2}dt^{2}+\frac{\left(\be dt+dr\right)^{2}}{1+E}+R^{2}d\Omega^{2},\label{metric}
\end{align}
where $\al(t,r)>0$ is the lapse function, $\be(t,r)$ is the radial component of the shift vector, $E(t,r)>-1$ reduces to the Lemaitre-Tolman-Bondi energy function in the uncharged dust limit \cite{lasky06a} and $d\Omega^{2}:=d\theta^{2}+\sin^{2}\theta d\phi^{2}$.  Furthermore, $R=R(t,r)$ is an arbitrary function whereby, without loss of generality, we set $R(t,0)=0$.  The normal vector in these coordinates was shown to be given by
\begin{align}
n^{\mu}=\frac{1}{\al}\left(-1,\,\be,\,0,\,0\right).
\end{align}
We were then able to show the equations governing the system were given by the following complicated set of equations.  Firstly, we defined a ``mass'' function in terms of the total energy density, $\rho$, and the total heat flux, $j$, by
\begin{align}
\frac{\da M}{\da r}:=4\pi\left(\rho\frac{\da R}{\da r}+j\Lie_{n}R\right)R^{2}.\label{massdef}
\end{align}    
Equations governing the evolution of $E$, $R$ and $M$ were then found to be given by
\begin{align}
\frac{\da R}{\da r}\frac{\Lie_{n}E}{2\left(1+E\right)}&=4\pi jR-\frac{1}{\al}\frac{\da R}{\da r}\frac{\da\be}{\da r}-\frac{\da}{\da r}\left(\Lie_{n}R\right),\label{LieEE}\\
\left(\Lie_{n}R\right)^{2}&=\frac{2M}{R}-1+\left(1+E\right)\left(\frac{\da R}{\da r}\right)^{2},\label{LieR}\\
\Lie_{n}M=-4\pi &R^{2}\left[\left(P-2\Pi\right)\Lie_{n}R+j\frac{\da R}{\da r}\left(1+E\right)\right],\label{LieM}\\
\frac{M}{R^{2}}+4\pi&\left(P-2\Pi\right)R=\frac{1+E}{\alpha}\frac{\da\al}{\da r}\frac{\da R}{\da r}-{\Lie_{n}}^{2}R,\label{Liesq}
\end{align}
where ${\Lie_{n}}^{2}R:=\Lie_{n}\left(\Lie_{n}R\right)$.  Furthermore, the final equation (\ref{Liesq}) is often trivially satisfied by the three preceding equations, however there are circumstances where this equation is required as we will see below.  There are two further equations coming from the conservation of energy momentum, equations (\ref{Lorentz}) and (\ref{energy}).  Using the metric (\ref{metric}) to express these equations they become long and aesthetically awkward and for this reason we suppress the independence of the EM and fluid contributions for the energy momentum tensor.  These two equations become respectively 
\begin{align}
&\frac{\mathcal{L}_{n}\left(j\sqrt{1+E}\right)}{\sqrt{1+E}}=2j\left[\frac{1}{2\left(1+E\right)}\mathcal{L}_{n}E+\frac{1}{\alpha}\frac{\partial\beta}{\partial r}-\frac{1}{R}\mathcal{L}_{n}R\right]
\nonumber\\
&-\frac{\left(\rho+P-2\Pi\right)}{\alpha}\frac{\partial\alpha}{\partial r}-\frac{\partial}{\partial r}\left(P-2\Pi\right)+\frac{6\Pi}{R}\frac{\partial R}{\partial r}.\label{Euler}\\
&\mathcal{L}_{n}\rho=\left(\rho+P-2\Pi\right)\left[\frac{1}{2\left(1+E\right)}\mathcal{L}_{n}E+\frac{1}{\alpha}\frac{\partial\beta}{\partial r}-\frac{2}{R}\mathcal{L}_{n}R\right]\nonumber\\
&-j\left(1+E\right)\left[\frac{1}{j}\frac{\partial j}{\partial r}+\frac{1}{2\left(1+E\right)}\frac{\partial E}{\partial r}+\frac{2}{R}\frac{\partial R}{\partial r}+\frac{2}{\alpha}\frac{\partial\alpha}{\partial 
r}\right]\nonumber\\
&-\frac{6\Pi}{R}\mathcal{L}_{n}R,\label{cont}
\end{align}
The system requires two further equations of state, for example where $P$ and $\Pi$ are expressed in terms of $\rho$ and $j$.  We will show in the coming sections that equations (\ref{Euler}) and (\ref{cont}) will reduce considerably for various subsystems of the general equations.  The two equations above are equivalent to the ones discussed at the end of section \ref{sec1}.  If we expand these equations in terms of the fluid components and the EM components of the energy momentum tensor, then indeed the term on the right hand side of equation (\ref{Lorentz}) appears in equation (\ref{Euler}).  

Summarily, the entire system of equations governing a spherically symmetric distribution of fluid is given by the line element (\ref{metric}), where the metric coefficients are related to the total contribution of the energy momentum tensor through equations (\ref{massdef}-\ref{cont}).  This total energy momentum tensor is decomposed into constituent fluid and EM contributions by equations (\ref{rhotot}-\ref{Pitot}).  In the next section we derive the electric and magnetic field components in spherical symmetry using the metric (\ref{metric}).  

\section{Charged Fluid}\label{sec3}
Spherical symmetry implies all vector fields can be written with only non-zero radial components.  Therefore, we make the following definitions
\begin{align}
E^{\mu}=&\left[0,\,\E(t,r),\,0,\,0\right],\\
B^{\mu}=&\left[0,\,\B(t,r),\,0,\,0\right],\\
i^{\mu}=&\left[0,\,i(t,r),\,0,\,0\right],
\end{align}
where $i$ is not to be confused with the complex number.  Maxwell's equations can now be expressed using the line element (\ref{metric}).  Beginning with equation (\ref{divB}), we find
\begin{align}
\frac{\da}{\da r}\left(\frac{\B R^{2}}{\sqrt{1+E}}\right)=&0.
\end{align}
Integrating this and using $R(t,0)=0$ implies the magnetic field in our frame vanishes everywhere 
\begin{align}
\B=0.\label{zeroB}
\end{align}
This is a direct consequence of the symmetry of the problem.  We note that we can still have a current in the spacetime, which would intuitively introduce a non-zero magnetic field.  However, as the current in the spacetime must be uniform across the two-spheres the magnetic field created everywhere will exactly cancel itself out.  Expressing equation (\ref{LieB}) in terms of the metric coefficients we find it vanishes identically.  Equations (\ref{divE}) and (\ref{LieE}) respectively become 
\begin{align}
\frac{\da}{\da r}\left(\frac{\E R^{2}}{\sqrt{1+E}}\right)=\frac{4\pi\ep R^{2}}{\sqrt{1+E}}&,\label{divE1}\\
\Lie_{n}\left(\frac{\E R^{2}}{\sqrt{1+E}}\right)=\frac{-4\pi iR^{2}}{\sqrt{1+E}}&.\label{LieE1}
\end{align}
Integrating equation (\ref{divE1}) we find the electric field is given by
\begin{align}
\E=\frac{\Q\sqrt{1+E}}{R^{2}},\label{EE}
\end{align}
where we have defined $\Q(t,r)$ such that
\begin{align}
\Q:=4\pi\int_{0}^{r}\frac{\ep R^{2}}{\sqrt{1+E}}dr.\label{Qdef}
\end{align}
This is interpreted as the total charge within a two-sphere of radius $r$ \cite{ori90a}.  We will see in the vacuum Einstein-Maxwell case\footnote{Throughout the article we refer to a vacuum in the sense that the fluid component of the energy momentum tensor vanishes, i.e. $\Tmat_{\mu\nu}=0$.  Of course in this case the total energy momentum tensor is still non-zero due to the EM component.}, it reduces to the charge of the RN solution (see section \ref{RN}).  Putting (\ref{EE}) back through the remaining Maxwell equation (\ref{LieE1}), we find 
\begin{align}
\Lie_{n}\Q=\frac{-4\pi iR^{2}}{\sqrt{1+E}}.\label{LieQ}
\end{align}   

Equations (\ref{zeroB}) and (\ref{EE}) give us equations of state for the EM contribution to the energy momentum tensor.  This is seen by evaluating equations (\ref{rhoem}-\ref{Piem}) with respect to $\Q$, from which it can be shown
\begin{align}
\rhoem=3\Pem=&\frac{3}{2}\Piem=\frac{\Q^{2}}{8\pi R^{4}},\label{EMEOS}\\
\jem=&0.\label{EMEOS1}
\end{align}
As mentioned, the $\jem$ component of the Maxwell energy momentum tensor is the Poynting flux.  This term being zero implies there is no transference of energy via the EM fields which is a direct result of the spherical symmetry.  As this term is zero, for the remainder of the article we drop subsequent superscripts on the $\jmat$ term.

Finally, equation (\ref{Lieep}) can be expressed in terms of the metric coefficients as
\begin{align}
\ep\Lie_{n}\left[\ln\left(\frac{\ep R^{2}}{\sqrt{1+E}}\right)\right]=-i\frac{\da}{\da r}\left[\ln\left(\frac{iR^{2}\al}{\sqrt{1+E}}\right)\right].\label{pp}
\end{align}

We have essentially shown that in the spherically symmetric Einstein-Maxwell system, the Maxwell equations can be integrated completely.  On the initial spacelike hypersurface we stipulate the charge density and the three-current density such that equation (\ref{pp}) is satisfied.  Furthermore, the Einstein constraint equations can be solved such that we know the geometry of the initial spacelike hypersurface, implying the metric coefficients are known initially.  Therefore, through equation (\ref{Qdef}) the charge per unit volume, $\Q$, is known on the initial hypersurface.  Equation (\ref{LieQ}) can then be integrated to give $\Q$ on subsequent spacelike hypersurfaces.   

Using (\ref{EMEOS}), (\ref{EMEOS1}) and the definition for the mass function (\ref{massdef}), we can show
\begin{align}
\frac{\da M}{\da r}=4\pi\left(\rhomat\frac{\da R}{\da r}+j\Lie_{n}R\right)R^{2}+\frac{\Q^{2}}{2R^{2}}\frac{\da R}{\da r},\label{massEM}
\end{align}
and equation (\ref{LieM}) becomes
\begin{align}
\Lie_{n}M=&-4\pi R^{2}\left[\left(\Pmat-2\Pimat\right)\Lie_{n}R+j\frac{\da R}{\da r}\left(1+E\right)\right]\nonumber\\
&+\frac{\Q^{2}}{2R^{2}}\Lie_{n}R.\label{LieMEM}
\end{align}

The following sections concentrate on reducing the general system of equations to specific examples and models.

\section{Exterior vacuum region}\label{RN}
A vacuum region is defined such that all fluid components of the energy momentum tensor vanish, i.e.
\begin{align}
\rhomat=\Pmat=\jmat=\Pimat=0.
\end{align}
Furthermore, if we are in vacuum there are no charge carrying particles and hence 
\begin{align}
\ep=i=0.
\end{align}
Using this we can integrate equations (\ref{Qdef}) and (\ref{LieQ}) to find 
\begin{align}
\Q=\Q_{0},
\end{align}
where $\Q_{0}$ is a constant equal to the total charge inside a sphere of certain radius being the boundary marking the vacuum region.  The mass function from equation (\ref{massEM}) can now be integrated and we find
\begin{align}
M=M_{0}-\frac{{\Q_{0}}^{2}}{2R},
\end{align}
where $M_{0}$ is a constant of integration.  

There are three remaining non-trivial equations, (\ref{LieEE}), (\ref{LieR}) and (\ref{Liesq}) which become respectively
\begin{align}
\frac{\Lie_{n}E}{2\left(1+E\right)}+\frac{1}{\al}\frac{\da\be}{\da r}=&-\frac{\da}{\da r}\left(\Lie_{n}R\right),\label{RN1}\\
\left(\Lie_{n}R\right)^{2}-\left(1+E\right)\left(\frac{\da R}{\da r}\right)^{2}=&-\left(1-\frac{2M_{0}}{R}+\frac{{\Q_{0}}^{2}}{R^{2}}\right),\label{RN2}\\
\frac{1+E}{\al}\frac{\da\al}{\da r}\frac{\da R}{\da r}-{\Lie_{n}}^{2}R=&\frac{M_{0}}{R^{2}}-\frac{{\Q_{0}}^{2}}{R^{3}}.\label{RN3}
\end{align}
The generalized Birkhoff's theorem states a spherically symmetric vacuum with an electric field is necessarily the RN spacetime \cite{misner73}.  Therefore, the metric (\ref{metric}) where the coefficients are solutions of equations (\ref{RN1}-\ref{RN3}) must be the RN spacetime.  The amount of generality corresponds to the vast generality for the interior regions of the spacetime.  Upon determining an interior, charged fluid filled region, one simply lets the metric coefficients be continuous across the boundary of the interface to determine the specific form of the RN region.  

It is trivial to see the equations governing the RN spacetime are a generalization of the equations governing the Schwarzschild spacetime given in \cite{lasky07}, whereby the Schwarzschild set are retrieved simply by setting $\Q_{0}=0$.

In general, the solution of equations (\ref{RN1}-\ref{RN3}) may be non-trivial.  However, we can pick out a handful of exact solutions by making {\it ad hoc} stipulations for some of the metric coefficients by using residual coordinate freedom available.  For example, by setting $\be=0$ and $R=r$, equation (\ref{RN2}) gives $E$ in terms of $M_{0}$ and $\Q_{0}$ and equation (\ref{RN1}) can subsequently be integrated for $\al$ and the metric is expressed in the usual RN coordinates.  Alternatively, Novikov-type coordinates are found by setting $\al=1$ and $\be=0$.  This implies equations (\ref{RN1}) and (\ref{RN2}) can be integrated such that the line element is
\begin{align}
dS^{2}=-dt^{2}+\frac{1}{\mathcal{F}_{1}}\left(\frac{\da R}{\da r}\right)^{2}dr^{2}+R^{2}d\Om^{2},
\end{align}
where $\mathcal{F}_{1}(r)$ is an arbitrary function of integration and $R$ is given implicitly by the solution of
\begin{align}
\int\frac{dR}{\sqrt{-1+2M_{0}/R-{\Q_{0}}^{2}/R^{2}+\mathcal{F}_{1}(r)}}=t+\mathcal{F}_{2}(r),\label{horror}
\end{align}
where $\mathcal{F}_{2}(r)$ is another function of integration.  The solution of this integral can be written down, however it is complicated and provides no new insight for the present discussion.  

Another set of coordinates can be obtained by assuming no temporal dependance in the line element.  Also letting $R=r$ implies the line element becomes  
\begin{align}
dS^{2}=&-\left(1-\frac{2M_{0}}{r}+\frac{{\Q_{0}}^{2}}{r^{2}}\right)dt^{2}\nonumber\\
&\pm2\sqrt{\frac{E+2M_{0}/r-{\Q_{0}}^{2}/r^{2}}{1+E}}dtdr\nonumber\\
&+\frac{dr^{2}}{1+E}+r^{2}d\Om^{2}.
\end{align}
This line element still contains a single free function of the radial coordinate, $E(r)$, which may be specified arbitrarily.  We note it is equivalent to the general line element expressed by \citet{patino01}, and if one chooses $\left(1+E\right)^{-1}=1-2M_{0}/r+{\Q_{0}}^{2}/r^{2}$, then without loss of generality, it results in the line element expressed by \citet{papapetrou74}.  

There is one more coordinate system that will prove relevant for section \ref{shell} on the collapse of infinitely thin shells and thus we introduce it here.  Generalized Painleve-Gullstrand type coordinates \cite{lasky06a} have canonical two-spheres and zero acceleration, i.e. $R=r$ and $\al=1$ respectively.  Putting this through equations (\ref{RN1}-\ref{RN3}) we find
\begin{align}
&\Lie_{n}E=0,\label{EPG}\\
\be^{2}=E&+\frac{2M_{0}}{r}-\frac{{\Q_{0}}^{2}}{r^{2}}.\label{betaPG}
\end{align}
Therefore, the only free function in the metric is the energy function $E$, which is a solution of (\ref{EPG}).  Furthermore, when taking the square root of (\ref{betaPG}) we can either choose a plus or minus sign, which give different regions of the RN solution on the Kruskal diagram (see for e.g. \cite{misner73}).  A trivial solution of (\ref{EPG}) is $E=0$, which are Painleve-Gullstrand coordinates for the RN spacetime \cite{painleve21,gullstrand22} (see \cite{poisson04} for a review of these coordinates in the Schwarzschild case).  For an expos\'e of more solutions of these equations and their properties see \cite{lasky07d}.

\section{Charged Vaidya}\label{Vad}
In \cite{lasky07} we used a coordinate transformation to go from the outgoing Vaidya metric to the $3+1$ line element, equation (\ref{metric}).  Here we generalize that result to include the charged Vaidya spacetime \cite{lindquist65,israel67,plebanski68,bonnor70} (also see \cite{hussain96,wang99} for more recent results).  We begin with the outgoing charged Vaidya metric 
\begin{align}
dS^{2}=-\left(1-\frac{2m}{R}\right)dv^{2}-2dvdR+R^{2}d\Om^{2},
\end{align}
where $v$ is an outgoing null coordinate, $(R,\theta,\phi)$ are spacelike coordinates and $m=m(v,r)$.  The energy momentum tensor may be decomposed with respect to two null vectors \cite{wang99}
\begin{align}
T_{\mu\nu}=\rhomat \ell_{\mu}\ell_{\nu}+\left(\om+\si\right)\left(\ell_{\mu}k_{\nu}+k_{\mu}\ell_{\nu}\right)+\si g_{\mu\nu},\label{VadT}
\end{align}
where the Einstein equations can be shown to satisfy
\begin{align}
\rhomat&=\frac{1}{4\pi R^{2}}\frac{\da m}{\da v},\label{rhomatvad}\\
\om&=\frac{1}{4\pi R^{2}}\frac{\da m}{\da R},\label{om}\\
\si&=\frac{1}{8\pi R}\frac{\da^{2}m}{\da R^{2}}.\label{si}
\end{align}
Furthermore, the vectors in (\ref{VadT}) are given in component form by
\begin{align}
\ell_{\mu}=&{\de_{\mu}}^{v},\\
k_{\mu}=&\frac{1}{2}\left(1-\frac{2m}{R}\right){\de_{\mu}}^{v}+{\de_{\mu}}^{R}.
\end{align}
One can see in (\ref{VadT}) if $\om=\si=0$, then the energy momentum tensor reduces to that of null dust and the spacetime reduces to Vaidya, whereby $\rhomat$ is the only non-zero component of the energy momentum tensor.  As $\rhomat$ in the Vaidya system is the only contribution to the energy momentum tensor which describes an uncharged null fluid, it can be shown $\om$ and $\si$ contribute to only the EM part of the energy momentum tensor in the charged Vaidya system \cite{wang99}.

We perform a coordinate transformation from the charged Vaidya line element to the non-null metric (\ref{metric}) such that $v=v(t,R)$ and $R=R(t,r)$.  This coordinate transformation is given by
\begin{align}
\frac{\da R}{\da t}=-\frac{\al m}{R}-\frac{\be\left(1-m/R\right)}{\sqrt{1+E}},&\,\,\,\frac{\da R}{\da r}=\frac{-\left(1-m/R\right)}{\sqrt{1+E}},\nonumber\\
\frac{\da v}{\da t}=-\al+\frac{\be}{\sqrt{1+E}},&\,\,\,\frac{\da v}{\da r}=\frac{1}{\sqrt{1+E}}.\label{coord}
\end{align}
We also write here the inverse coordinate transformation as this will be necessary for the later derivations
\begin{align}
\frac{\da t}{\da v}=\frac{-1}{\al}\left(1-\frac{m}{R}\right),&\,\,\,\,\,\frac{\da t}{\da R}=\frac{-1}{\al},\\
\frac{\da r}{\da v}=\frac{\be}{\al}\left(1-\frac{m}{R}\right)+\frac{m}{R}\sqrt{1+E},&\,\,\,\,\,\frac{\da r}{\da R}=\frac{\be}{\al}-\sqrt{1+E}.
\end{align}

One can see the coordinate transformation (\ref{coord}) implies the relation
\begin{align}
\Lie_{n}R=-\frac{m}{R}.\label{vadLieR}
\end{align}
Putting this through equation (\ref{LieR}) we find the result that the mass defined in equation (\ref{massdef}) reduces to the mass of the charged Vaidya system 
\begin{align}
M=m.
\end{align} 
Now, for solutions of the differential equations defining the coordinate transformation to exist and hence for the coordinate transformation to be valid, the integrability conditions of the coordinate transformation must be satisfied.  These integrability conditions provide equations constraining the form of the metric coefficients.  After some algebra we find these to be
\begin{align}
\frac{\sqrt{1+E}}{\al}\frac{\da\al}{\da r}=&\frac{\Lie_{n}E}{2\left(1+E\right)}+\frac{1}{\al}\frac{\da\be}{\da r}\label{int1}\\
=&\frac{-1}{R}\left(\Lie_{n}M+\sqrt{1+E}\frac{\da M}{\da r}\right)-\frac{M}{R^{2}}.\label{int2}
\end{align}
Putting the integrability conditions through equations (\ref{LieEE}) and (\ref{Liesq}) we find after some algebra
\begin{align}
\Lie_{n}M&+\sqrt{1+E}\frac{\da M}{\da r}\nonumber\\
&=4\pi R^{2}\left(\Pmat-2\Pimat-j\sqrt{1+E}\right)-\frac{\Q^{2}}{2R^{2}}.\label{LieM+daMdar}
\end{align}
Furthermore, putting the integrability equations and the coordinate transformation through equations (\ref{massdef}) and (\ref{LieM}) and substituting through equation (\ref{LieM+daMdar}), we can show the fluid variables satisfy a general equation of state
\begin{align}
\rhomat+\Pmat-2\Pimat-2j\sqrt{1+E}=0\label{genEOS}
\end{align}

Now, by expanding the derivative with respect to the null coordinate $v$ in equation (\ref{rhomatvad}) with respect to the temporal and radial coordinates, $t$ and $r$, one can show
\begin{align}
\frac{\da M}{\da v}=-4\pi R^{2} j\sqrt{1+E}.
\end{align}
Hence, direct comparison with equation (\ref{rhomatvad}) implies
\begin{align}
\rhomat=j\sqrt{1+E}
\end{align}
Doing the same for equation (\ref{om}) and (\ref{si}) we find respectively
\begin{align}
\om=&\frac{\Q^{2}}{8\pi R^{4}}\equiv\rhoem,\\
\si=&\frac{\Q}{2R}\left(\ep-\frac{i}{\sqrt{1+E}} \right)-\frac{\Q^{2}}{8\pi R^{4}}.\label{si2}
\end{align}

Now, substituting the conditions for the coordinate transformations and equations (\ref{genEOS}-\ref{si2}) into the conservation of energy momentum equations, (\ref{Euler}) and (\ref{cont}), after much algebra we can show these reduce to the simple relation
\begin{align}
\si+\rhoem=\rhomat+3\Pimat.
\end{align}

The uncharged Vaidya system had four energy momentum variables with three equations of state implying we could express all energy momentum variables in terms of just one of the variables.  However, the charged Vaidya system has six energy momentum variables, $\rhomat$, $\Pmat$, $\Pimat$, $j$, $\rhoem$ and $\si$.  However, we still only have three equations of state which can be summarized from the above as
\begin{align}
\rhomat=&j\sqrt{1+E}\\
=&\Pmat-2\Pimat\\
=&\rhoem+\si-3\Pimat. 
\end{align}
It is trivial to show when $\rhoem=\si=0$, the system reduces to the uncharged Vaidya in non-null coordinates \cite{lasky07}.  The extra generality in this system is attributed to the extra generality associated with the charge distributions and currents throughout the spacetime.  

Two fmore equations from this system will prove useful in the further analysis.  Firstly, putting the equations of state through equation (\ref{LieM+daMdar}) we find
\begin{align}
\Lie_{n}M+\sqrt{1+E}\frac{\da M}{\da r}=-\frac{\Q^{2}}{2R^{2}}.\label{dododo}
\end{align}
Secondly, putting this equation back through the integrability condition, equation (\ref{int2}), implies
\begin{align}
\frac{1}{\al}\frac{\da\al}{\da r}=&\frac{-1}{R^{2}\sqrt{1+E}}\left({M}-\frac{\Q^{2}}{2R}\right).\label{dadada}
\end{align}

Summarily, the system of equations governing the charged Vaidya spacetime in non-null coordinates is the line element (\ref{metric}) where the metric coefficients are given by (\ref{vadLieR}), (\ref{int1}), (\ref{dododo}), and (\ref{dadada}).  One can imagine establishing an initial value problem such that an initial spacelike hypersurface is established with an interior charged fluid ball reaching some finite radius and a vacuum RN region extending beyond this radius.  The system then evolves and the fluid ball is allowed to radiate null radiation such that a region of charged Vaidya as described above develops between the fluid ball and the vacuum.  Taking any subsequent $t$ equals constant slice of the spacetime will result in a region of a charged fluid ball with charged Vaidya extending beyond this and RN reaching out to radial infinity.  In this way causality is provided for the charged Vaidya spacetime within a realistic physical model.  Of course the equations described herein are difficult to solve analytically, as such future work will provide a numerical realization of this paradigm.

\section{Thin Charged shell}\label{shell} 
Consider an infinitely thin shell of charged matter falling into an RN black hole.  As the shell is infinitely thin, there are no internal forces between particles other than the Coulomb forces implying the fluid is dust, i.e. $\Pmat=\Pimat=j=0$.  Therefore, the only non-zero contribution to the fluid component of the energy momentum tensor is from the energy density, which is given by
\begin{align}
\rhomat=\rhomat_{0}\de(r)+\rhomat_{\star}\de\left(r-r_{\star}\right).
\end{align}
Here, $\de$ is the Dirac delta, $\rhomat_{0}\in\real$ is the energy density of the initial RN black hole situated at $r=0$ and $\rhomat_{\star}\in\real$ is the energy density of the shell, which is located at $r=r_{\star}(t)$.  Furthermore, the dust particles of the collapsing shell carry charge and hence the charge density is given by
\begin{align}
\ep=\ep_{0}\de(r)+\ep_{\star}\de\left(r-r_{\star}\right),
\end{align}
where $\ep_{0}\in\real$ is the charge density of the point mass at $r=0$ and $\ep_{\star}\in\real$ is the charge density of the shell.  

The energy momentum tensor here is described by two variables, $\rhomat$ and $\ep$.  As there are four degrees of freedom in the metric ($\al$, $\be$, $E$ and $R$) and only two degrees of freedom in the energy momentum tensor, we have residual coordinate freedom in the line element (see the discussion in \cite{lasky07}).  Hence, without loss of generality we use canonical two-spheres in the following derivations, i.e.
\begin{align}
R=r.
\end{align}  
We will use the remaining coordinate freedom associated with the spacetime implicitly at a later stage of the derivation to further simplify the mathematics.  Equation (\ref{Qdef}) can now be integrated to give
\begin{align}
\Q=\Q_{0}+\Q_{\star}\U\left(r-r_{\star}\right),
\end{align}
where $\Q_{0},\Q_{\star}\in\real$ are constants of integration which represent the charge at $r=0$ and of the shell respectively and $\U$ is the Heaviside step function.  Integrating the definition of the mass function (\ref{massdef}) we find
\begin{align}
M=&M_{0}-\frac{{\Q_{0}}^{2}}{2r}+\left[M_{\star}-\frac{\Q_{\star}}{2r}\left(2\Q_{0}+\Q_{\star}\right)\right]\U\left(r-r_{\star}\right)\nonumber\\
=&\left\{\begin{array}{cc}
M_{0}-\frac{{\Q_{0}}^{2}}{2r} & r<r_{\star}\\
 & \\
\left(M_{0}+M_{\star}\right)-\frac{\left(\Q_{0}+\Q_{\star}\right)^{2}}{2r} & r\ge r_{\star}\end{array}\right..\label{shellmass}
\end{align}   
Here, $M_{0}$ is the mass of the initial Schwarzschild black hole and $M_{\star}$ is another constant of integration which is the mass of the shell.  Equation (\ref{shellmass}) gives the expected result that for $r<r_{\star}$ the spacetime is the RN spacetime with mass and charge contributions being given by $M_{0}$ and $\Q_{0}$ respectively.  The region for which $r\ge r_{\star}$ is also the RN spacetime, however the mass and charge contributions are given by $M_{0}+M_{\star}$ and $\Q_{0}+\Q_{\star}$ respectively.  This further implies that if the charge at $r=0$ and at the shell are of equal but opposite sign (for example protons and electrons), then the exterior region of the spacetime, $r\ge r_{\star}$, is given simply as the Schwarzschild spacetime with total mass $M_{0}+M_{\star}$.  That is, exterior to the shell in this scenario one can not feel the effects of the charges (this is discussed more in section \ref{exschw}). 

Now, putting the derived form of the mass function through equation (\ref{LieM}), we find after some algebra
\begin{align}
\left[M_{0}-\frac{\Q_{\star}}{2r}\left(2\Q_{0}+\Q_{\star}\right)\right]\Lie_{n}\left[\U\left(r-r_{\star}\right)\right]=0.
\end{align}
Hence, provided $M_{0}\neq\Q_{\star}\left(2\Q_{0}+\Q_{\star}\right)/2r$, we find
\begin{align}
\left(\frac{dr_{\star}}{dt}+\be\right)\de\left(r-r_{\star}\right)=0.
\end{align}
This is trivially satisfied for all $r\neq r_{\star}$ and provides no new information.  However, for $r=r_{\star}$, we see
\begin{align}
\frac{dr_{\star}}{dt}=-\be_{\star},\label{drstar}
\end{align}
where $\be_{\star}:=\be\left[t,r_{\star}(t)\right]$ is the shift function evaluated on the shell of matter.  Equation (\ref{drstar}) implies the observer is moving with the velocity of the shell of matter.  

Evaluating the two conservation of energy momentum equations, we find (\ref{cont}) is trivially satisfied and (\ref{Euler}) becomes after some algebra
\begin{align}
\frac{\rhomat_{0}\de(r)+\rhomat_{\star}\de\left(r-r_{\star}\right)}{\alpha}\frac{\da\al}{\da r}=\frac{\Q_{\star}}{8\pi r^{4}}\left(2\Q_{0}+\Q_{\star}\right)\de\left(r-r_{\star}\right).\label{al}
\end{align}
This equation can be integrated directly and one recovers two boundary conditions for the lapse function.  The first of these boundary conditions is at $r=0$, however there is a physical singularity at this point implying the spacetime is only well defined for $r>0$.  Therefore this first boundary condition is physically irrelevant for the system in study.  The second boundary condition derived from the integration of equation (\ref{al}) is at the boundary of the shell of charged matter  
\begin{align}
\left.\frac{\da}{\da r}\left(\ln\al\right)\right|_{r_{\star}}=\frac{k}{r_{\star}^{4}},\label{bound2}
\end{align}
where we have defined the constant 
\begin{align}
k:=\frac{\Q_{\star}\left(2\Q_{0}+\Q_{\star}\right)}{8\pi\rho_{\star}}.\label{kdef}
\end{align}
This boundary condition implies the lapse is either steeply increasing or decreasing across the boundary of the charged shell depending on the sign of the respective charges.  Furthermore, we see that if the charge of the shell, $Q_{\star}$, and the charge of the point mass, $Q_{0}$, are of the same sign then the gradient of the lapse at $r_{\star}$ is necessarily increasing.  The case where $k=0$ is a greatly simplified scenario as it implies $\al$ can be set to unity throughout the spacetime (see section \ref{zerok}).     


The lapse and energy functions are related through equation (\ref{LieE}), which reduces to
\begin{align}   
\frac{\Lie_{n}E}{2\left(1+E\right)}=\frac{\be}{\al}\frac{\da}{\da r}\left(\ln\al\right).\label{de}
\end{align}
This equation can be evaluated on the shell and we find by using equation (\ref{drstar}) and the chain rule that the left hand side simply becomes an ordinary timelike derivative of the energy function along the shell.  Furthermore, by using the boundary condition (\ref{bound2}), the entire equation can be re-expressed as
\begin{align}
\frac{d}{dt}\left(\ln\sqrt{1+E_{\star}}\right)=\frac{k}{r_{\star}^{4}}\frac{dr_{\star}}{dt},
\end{align}
where $E_{\star}(t):=E\left[t,r_{\star}(t)\right]$.  Upon integrating this equation with respect to $t$ we introduce an arbitrary constant which relates the initial energy function on the shell to the initial radius of the shell.  We can arbitrarily stipulate the initial energy function on the shell such that this arbitrary constant equals unity without loss of generality.  It is subsequently found that the energy function on the shell is given by
\begin{align}
1+E_{\star}=\exp\left(\frac{-2k}{3r_{\star}^{3}}\right).\label{Estar}
\end{align}
As mentioned, $E_{\star}$ is the value of the energy function along the shell as it evolves through the spacetime.  Therefore, in keeping consistent with the current paradigm, we choose to let $E(t,r)$ be a continuous function throughout the spacetime such that for $r=r_{\star}$ the energy function reduces to the form given in equation (\ref{Estar}).  That is we stipulate $E(t,r)$ for the entire spacetime to be
\begin{align}
1+E=\exp\left(\frac{-2k}{3r^{3}}\right).\label{E}
\end{align}
We note here that this actually implies the energy function is independent of the temporal coordinate.  In doing this we have lost some generality in the solution as to keep the solution as general as possible would have required the introduction of extra arbitrary functions whilst still ensuring the form of $E$ reduces to that of $E_{\star}$ at $r=r_{\star}$.  However, the choice we have made above allows the solution of the lapse function to be solved with ease.  We will show that this loss in generality is equivalent in the non-charged case to choosing the marginally bound solution, i.e. $E=0$\footnote{The marginally bound solution does not exist in the charged case as if we let $E=0$ in equation (\ref{de}) then in general the boundary condition (\ref{bound2}) is not satisfied.}.  Furthermore, this is equivalent to choosing the initial velocity of the shell on the initial hypersurface. 

By putting the form of the energy function, equation (\ref{E}), back through equation (\ref{de}) we find the lapse function is also independent of the temporal coordinate.  This simplifies the mathematics as it implies the shift function can be cancelled and the equation can be directly integrated for the lapse function, which we find becomes
\begin{align}
\al=\exp\left(\frac{-k}{3r^{3}}\right).
\end{align}
Here, an arbitrary constant of integration has been set to unity by re-scaling the temporal coordinate.  It is trivial to show that this form of the lapse function satisfies the boundary condition given by equation (\ref{bound2}).

Finally, equation (\ref{LieR}) with $R=r$ implies the shift function can be expressed in terms of the lapse, energy and mass functions according to
\begin{align}
\be^{2}=\al^{2}\left(E+\frac{2M}{r}\right).\label{be}
\end{align}
Taking the positive square root of this equation gives a model which is collapsing and the negative root gives an expanding shell \cite{lasky06a}.  We now know all the metric coefficients for the spacetime and hence the system is almost entirely determined.  The only remaining piece of information is to determine the position of the shell as a function of the temporal coordinate.  We note that equation (\ref{be}) can be evaluated on the shell and this gives a first order, non-linear ordinary differential equation for the position of the shell as a function of time
\begin{align}  
\frac{dr_{\star}}{dt}=&-e^{-k/3r_{\star}^{3}}\sqrt{e^{-2k/3r_{\star}^{3}}-\left[1-\frac{2M\left(t,r_{\star}\right)}{r_{\star}}\right]},\label{rstarde}
\end{align}
where
\begin{align}
M(t,r_{\star})=\left(M_{0}+M_{\star}\right)-\frac{\left(\Q_{0}+\Q_{\star}\right)^{2}}{2r_{\star}}.
\end{align}
This differential equation governs the dynamics of the system in question, with different charge species and strengths giving different evolutions for the shell.  It is interesting to note that for $r_{\star}<<1$ the exponential terms become extremely small (providing $k\neq0$) and one finds $dr_{\star}/dt\approx0$, implying $r_{\star}\,\in\,\real$.  Hence, any infalling shell in this coordinate system will take infinitely long to reach $r=0$.     

For completeness we write down the line element here
\begin{align}
dS^{2}=&-\left(1-\frac{2M}{r}\right)dt^{2}\nonumber\\
&+2\sqrt{1-\exp\left(\frac{2k}{3r^{3}}\right)\left(1-\frac{2M}{r}\right)}\,dtdr\nonumber\\
&+\exp\left(\frac{2k}{3r^{3}}\right)dr^{2}+r^{2}d\Om^{2},\label{nice}
\end{align}
where $M(t,r)$ is given by equation (\ref{shellmass}).  One can see that the only function left to be determined in the line element is $r_{\star}$ which appears implicitly through the mass function and is the solution of equation (\ref{rstarde}).  

Either side of the infinitely thin shell, the spacetime is vacuum and must necessarily be given by the RN spacetime.  Indeed evaluating the Einstein tensor for the line element given in (\ref{nice}) either side of the shell gives an energy momentum tensor in accordance with the RN spacetime.  Hence, the line element given above with $M=\bar{M}-\bar{\Q}^{2}/2r$ (where $\bar{M}$ is a constant representing either $M_{0}$ or $M_{0}+M_{\star}$ and $\bar{\Q}$ is either  $\Q_{0}$ or $\Q_{0}+\Q_{\star}$) describes the RN spacetime.  One can also look at the asymptotics of the spacetime and find that the metric tends to the Minkowski metric as $r\rightarrow\infty$.  This is consistent with the earlier description that this is equivalent to the marginally bound case with charge as the metric tending to the Minkowski metric implies the observer at infinity has zero energy.  It is interesting to note that this spacetime is a specific form of that given by \citep{patino01}, in particular, taking equation (13) from \cite{patino01} with $c(r)=\exp(2k/3r^{3})$.      

\subsection{Apparent Horizon}
An important facet of the discussion of specific solutions of equation (\ref{rstarde}) and associated forms of the line element (\ref{nice}) is the evolution of the apparent horizon.  It is well established that there are up to two horizons in the RN spacetime, which we denote as $\rAH$.  In \cite{lasky06b} we derived the apparent horizon for any spacetime with line element given by equation (\ref{metric}) with $R=r$ to be
\begin{align}
\rAH(t)=2M[t,\rAH(t)],
\end{align}
Substituting equation (\ref{shellmass}) into the apparent horizon and solving for $\rAH$ in the two regions we find 
\begin{align}
\rAH(t)=M_{0}\pm\sqrt{{M_{0}}^{2}-{\Q_{0}}^{2}},
\end{align}
for $\rAH<r_{\star}$ and
\begin{align}
\rAH(t)=M_{0}+M_{\star}\pm\sqrt{\left(M_{0}+M_{\star}\right)^{2}-\left(\Q_{0}+\Q_{\star}\right)^{2}},
\end{align}
for $\rAH\ge r_{\star}$.  These are the well known forms of the apparent horizon \cite{misner73} and we see there is a jump discontinuity across the thin shell of matter.

We note here that the RN spacetime can be broken into four distinct subcases which has relevance for the apparent horizon and we will also see has relevance for the collapsing shells.  Consider for the moment the RN spacetime with mass and charge $\bar{M}$ and $\bar{\Q}$ respectively. 
The four cases are as such
\begin{enumerate}
\item When $\bar{\Q}=0$ the spacetime is simply Schwarzschild.  In this case $r_{-}=0$ and $r_{+}=2\bar{M}$.
\item When $0<\bar{\Q}<\bar{M}$ the interior horizon is $0<r_{-}<\bar{M}$ and the exterior horizon is $\bar{M}<r_{+}<2\bar{M}$.
\item $\bar{\Q}=\bar{M}$ is known as an extremal black hole and we note that the interior and exterior horizons coincide, i.e. $r_{-}=r_{+}=\bar{M}$.
\item A model of a classical charged particle is often considered with $\bar{\Q}>\bar{M}$.  In this case we notice that the discriminant in the apparent horizon becomes negative and hence neither apparent horizon exist.  This is therefore a naked singularity as the point mass at $r=0$ is visible by an observer at infinity.   
\end{enumerate}  

\subsection{Shell proper motion}
One can study the motion of the shell in a coordinate invariant way by looking at the proper motion of the charged shell.  The proper time is found by looking at the geodesic equation coming from the metric.  Spherical symmetry implies the motion of the shell is independent of the angular coordinates and one can rearrange the metric to read
\begin{align}
\left(\frac{d\tau}{dt}\right)^{2}=\left[\al^{2}-\frac{\be^{2}}{1+E}-\frac{2\be}{1+E}\frac{dr}{dt}-\frac{1}{1+E}\left(\frac{dr}{dt}\right)^{2}\right],
\end{align}
where $d\tau^{2}=-dS^{2}$ and we have reverted back to the $3+1$ notation for brevity in the expressions.  We are only concerned with the motion of the shell and hence only interested in the relation of the proper time and coordinate time along the shell.  Therefore, we can substitute equation (\ref{drstar}) into the above, invert and take the square root to find the simple relation between the coordinate and proper times evaluated along the shell
\begin{align}
\left.\frac{dt}{d\tau}\right|_{r_{\star}}=&\frac{1}{\al_{\star}}\nonumber\\
=&\exp\left(\frac{k}{3r_{\star}^{3}}\right).
\end{align}
We note that when $k=0$ the coordinate time and proper time along the shell are equivalent.  

The proper motion of the shell can now be calculated from equation (\ref{rstarde}) by substituting the above and we find
\begin{align}
\frac{dr_{\star}}{d\tau}=-\sqrt{e^{-2k/3r_{\star}^{3}}-\left[1-\frac{2M\left(\tau,r_{\star}\right)}{r_{\star}}\right]}.\label{rstarde2}
\end{align}   
One can gain much information about the motion of the shell by looking at its acceleration.  This gives insight into the relativistic ``forces'' in the spacetime.  Differentiating with respect to the proper time we find
\begin{align}
a_{\star}=\frac{k}{r_{\star}^{4}}e^{-2k/3r_{\star}^{3}}-\frac{\M}{r_{\star}^{2}}+\frac{\left(\Q_{0}+\Q_{\star}\right)^{2}}{r_{\star}^{3}},\label{accel}
\end{align}
where $a_{\star}(\tau):=d^{2}r_{\star}/d\tau^{2}$ is the proper acceleration of the shell and we have defined
\begin{align}
\M:=M_{0}+M_{\star},
\end{align}
for convenience.  It is of interest to explore the regimes where the acceleration is positive and negative as this gives us insight into the nature of the motion of the shell.  However, the expressions for the velocity and acceleration of the shell are difficult to analyse in general, so we revert to specific examples to highlight different physical aspects.

\subsection{Shell Examples}
\subsubsection{Zero Charge}\label{zero}
As a first example we reduce the equations to those of non-charged dust.  It is interesting to see that there is still a difference between a non-charged shell of matter falling onto a Schwarzschild black hole as compared to an RN black hole.  This is exhibited in the equation governing the evolution of the shell (\ref{rstarde}) as the Schwarzschild black hole has $\Q_{0}=0$ whereas it is nonzero for the RN black hole.  Despite the non-charged shell not interacting through EM forces in the RN case, the geometry of an RN black hole is significantly different to the Schwarzschild black hole, which is exhibited through the different motion of the shell.

For this example we look at the evolution of the non-charged shell simply falling onto a Schwarzschild black hole, i.e. $\Q_{0}=\Q_{\star}=0$.  This implies from its definition that the constant $k$ vanishes, which implies the coordinate time and the proper time are equivalent.  Furthermore, equation (\ref{rstarde2}) greatly simplifies and can be integrated to give 
\begin{align}
r_{\star}^{3/2}-r_{\star}(0)^{3/2}=-\frac{3\tau}{2}\sqrt{2\M}.\label{geo}
\end{align}
This equation is exactly that of a geodesic for a zero energy particle and is the model elucidated by \citet{adler05}.  It is trivial to see from the above that as proper time evolves, $r_{\star}$ moves closer to $r=0$ and will eventually collide with the singularity, as expected.  We can calculate the acceleration of the shell here such that we can compare with later results and we find
\begin{align}
a_{\star}=-\frac{\M}{r_{\star}^{2}}.
\end{align} 
As the mass is positive definite the acceleration of the shell in this case must always be negative.  Therefore, as the only influences on the shell of particles are gravitational, we discern that a negative acceleration corresponds to an attractive ``force''.\footnote{This is actually due to the curvature of the spacetime, however we loosely use the term ``force'' as we will be comparing the effects due to gravitation with that of the Coulomb force.}

\subsubsection{Zero $k$}\label{zerok}
Analytic solutions of (\ref{rstarde2}) can also be found when $k=0$.  We see this is the case when either 
\begin{align}
\Q_{\star}=0\qquad\textrm{and}\qquad\Q_{0}\neq0,
\end{align}
or
\begin{align}
2\Q_{0}=-\Q_{\star}\label{q0qstar}
\end{align}
In fact, the dynamics of the entire spacetime are indistinguishable for these two charge distributions.  Thus an uncharged shell of matter falling onto an RN black hole is equivalent to the charges being given by equation (\ref{q0qstar}).  For the remainder of this section we shall discuss the $\Q_{\star}=0$ case, however we note this is equally applicable to the $2\Q_{0}=-\Q_{\star}$ case.

Putting $k=0$ into the line element we find it becomes
\begin{align}
dS^{2}=&-\left(1-\frac{2M}{r}\right)dt^{2}+2\sqrt{\frac{2M}{r}}\,dtdr+dr^{2}+r^{2}d\Om^{2},
\end{align}
where $M$ is still given by (\ref{shellmass}) with $\Q_{\star}=0$.  This line element is the RN spacetime in Painleve-Gullstrand type coordinates both interior and exterior to the shell.  While the Painleve-Gullstrand coordinates for the Schwarzschild spacetime cover the regions from $r=0$ to infinity, the equivalent version for the RN spacetime do not.  One can see this by looking at the $g_{tr}$ component of the line element, which reads 
\begin{align}
g_{tr}=\sqrt{\frac{2\bar{M}}{r}-\frac{\bar{\Q}^{2}}{r^{2}}}.
\end{align}
We see that this coefficient becomes complex for
\begin{align}
r<r_{c}:=\frac{\bar{\Q}^{2}}{2\bar{M}},\label{rc}
\end{align}
where we have defined the critical radius of the spacetime, $r_{c}$.  That is, the coordinates are invalid for $r<r_{c}$ and therefore they have a range given by $r\in\left[\left.r_{c},\infty\right)\right.$.  One can compare this to the position of the horizon in the RN spacetime.  In particular taking two extreme limits of the RN black hole, namely $\bar{\Q}=0$ and $\bar{\Q}=\bar{M}$.  Firstly, we find when $\bar{\Q}=0$ the critical radius also vanishes and hence is not an issue.  At the other limit to consider is that of an extremal black hole where $\bar{Q}=\bar{M}$ one finds $\rAH=\bar{M}>\bar{M}/2=r_{c}$.  Hence we deduce for all RN black holes such that $0\le\bar{\Q}\le\bar{M}$ the critical radius is inside the apparent horizon.  For black holes with $\bar{Q}>\bar{M}$ the apparent horizon does not exist and the critical radius is therefore visible to an external observer.     

Possibly one of the most interesting things about this line element is the potential link with microscopic physics.  The critical radius defined in (\ref{rc}) is actually the Compton radius which describes the ``electron radius'' (see for example \cite{feynman64}).  Furthermore, we know that the apparent horizon, $r_{+}$, for an electron is necessarily inside this radius, $r_{+}<r_{c}$.  Hence, this line element is a way of classically describing the gravitational and electromagnetic field of an electron without having to consider the complicated interior region or by modelling it as a point-mass.    

Reverting back to the collapsing shell within this spacetime, the evolution of the shell becomes
\begin{align}
\frac{dr_{\star}}{d\tau}=-\sqrt{\frac{2\M}{r_{\star}}-\frac{\Q_{0}^{2}}{r_{\star}^{2}}}.\label{tjnbw}
\end{align}
It is interesting that the sign of $\Q_{0}$ does not matter to the evolution of this system.  That is, changing the species of $\Q_{0}$ from negative charge to positive charge does not effect the system provided the sign of $\Q_{\star}$ also changes such that equation (\ref{q0qstar}) is satisfied.  It can also be shown from equation (\ref{tjnbw}) that $dr_{\star}/d\tau$ vanishes when
\begin{align}
r_{\star}(t)=r_{c}.
\end{align}
Furthermore, when $r_{\star}<r_{c}$ the term in the square root sign of equation (\ref{tjnbw}) becomes negative and hence there are no real solutions of this equation. 

Integrating equation (\ref{tjnbw}) gives
\begin{align}
\tau+\mathcal{C}=\frac{-1}{3\M^{2}}\sqrt{2\M r_{\star}-\Q_{0}^{2}}\left(\Q_{0}^{2}+\M r_{\star}\right),
\end{align}
where $\mathcal{C}$ is an arbitrary constant of integration.  One can see when $\Q_{0}=0$ this equation reduces to equation (\ref{geo}) describing the non-charged shell as expected.  By putting $r_{\star}=r_{c}$ into the above solution, one can show that the critical proper time, $\tau_{c}$, defined to be the proper time at which the shell crosses the critical radius is given by
\begin{align}
\tau_{c}=\frac{1}{3\M^{2}}\sqrt{2\M r_{\star}(0)-\Q_{0}^{2}}\left[\Q_{0}^{2}+\M r_{\star}(0)\right],
\end{align}
which is positive definite.  

We can highlight specific features of the collapsing shell by looking at its acceleration throughout the spacetime as this is proportional to the ``forces'' acting on the shell.  Evaluating equation (\ref{accel}) with $\Q_{\star}=0$ we find 
\begin{align}
a_{\star}=-\frac{1}{r_{\star}^{2}}\left(\M-\frac{\Q_{0}^{2}}{r_{\star}}\right).
\end{align}
Therefore we find
\begin{align}
a_{\star}\le0\qquad\Longleftrightarrow\qquad r_{\star}\ge\frac{\Q_{0}^{2}}{\M}:=r_{a}.
\end{align}
In direct comparison with the uncharged case (section \ref{zero}), a negative acceleration corresponds to an attractive force on the shell.  However, we see that for $r_{c}\le r_{\star}<\Q_{0}^{2}/\M$ the acceleration of the shell is positive and hence the force on the shell is repulsive.  The value $r_{a}$ is therefore the point at which the acceleration changes sign.  As $\Q_{\star}=0$ in this case, the infalling shell is simply dust and hence there are no internal forces between the particles in the shell.  Furthermore, the shell is uncharged implying there is no Coulomb force from the charged point mass at $r=0$ acting on the shell.  We saw that this repulsive term was not present in the case of an uncharged shell falling onto a Schwarzschild black hole.  Therefore this repulsion is simply a manifestation of the geometry of the RN spacetime associated with the presence of the extra term in the energy momentum tensor associated with the charge.    

We saw that the acceleration of the shell is positive for a finite range of the radius, however the position of this range as compared with the apparent horizon is interesting.  Again we divide the spacetime into four distinct subcases;
\begin{enumerate}
\item When $\Q_{0}=0$ we find $r_{a}=0$ and hence the force is always attractive (this is the uncharged case discussed in section \ref{zero}). 
\item When $\Q_{0}<\M$ we can show that $r_{-}<r_{a}<r_{+}$.  That is, the position at which the acceleration changes sign is hidden behind the apparent horizon of the RN spacetime.  Therefore this effect is not observable from an external observer.
\item When $\Q_{0}=\M$ the two apparent horizons and the position of the change in acceleration also align, that is $r_{\pm}=r_{a}=\M$.  
\item The interesting case is again where $\Q_{0}>\M$ implying the apparent horizons do not exist.  In this case we find $r_{a}>\M$, implying this effect is visible by an external observer.  
\end{enumerate}

\subsubsection{Exterior Schwarzschild}\label{exschw}
As a next example we look at the case where
\begin{align}
\Q_{0}=-\Q_{\star}.
\end{align}
This example is interesting as the mass function for the exterior region, $r>r_{\star}$, is simply $M=M_{0}+M_{\star}$, implying this exterior region is Schwarzschild.  The charges in this case are equal but of opposite sign, which are therefore known to possess an attractive Coulomb force.  We find from equation (\ref{kdef}) that $k<0$ and the evolution of the shell is given by
\begin{align}
\frac{dr_{\star}}{d\tau}=-\sqrt{\exp\left(\frac{\Q_{0}^{2}}{12\pi\rho_{\star}r_{\star}^{3}}\right)-1+\frac{2\M}{r_{\star}}}.\label{equation}
\end{align}
Furthermore, equation (\ref{accel}) can be shown to be negative definite implying the acceleration is always negative, hence the force is always attractive.  

Of interest in this model is that an observer external to the collapsing shell, i.e. sitting at some $r>r_{\star}$ is in a region of spacetime equivalent to the Schwarzschild spacetime.  However the observer can view the motion of the shell and infer the existence of the charges as the motion of the shell is significantly different to the uncharged case.


\section{Static Equations}\label{static}
Static charged fluid spheres have been studied in great detail, especially perfect fluid spheres in diagonal coordinates \cite{ivanov02}.  The method we have utilized herein can also find static solutions by finding a timelike Killing vector that is hypersurface orthogonal which implies the spacetime is static \cite{wald84} and using this vector to reduce the general equations.  Such a Killing vector is given by \cite{lasky07}
\begin{align}
N^{\mu}=\frac{\da}{\da t}-\beta\frac{\da}{\da r}.
\end{align}
Furthermore, solving Killing's equations with this vector field gives equations which further constrain the metric coefficients such that the spacetime is static.  These equations are given by
\begin{align}
\Lie_{n}R=&0,\label{static1}\\
\frac{\Lie_{n}E}{2\left(1+E\right)}=&\frac{-1}{\al}\frac{\da\be}{\da r},\label{static2}\\
\Lie_{n}\al=&0.\label{static3}
\end{align}
Equations (\ref{static1}) and (\ref{static2}) together with equation (\ref{LieEE}) imply that a static, spherically symmetric Einstein-Maxwell spacetime necessarily has vanishing heat flux
\begin{align}
j=0.
\end{align}
Putting the above conditions through the remaining non-trivial fluid and EM equations, (\ref{LieR}), (\ref{Liesq}), (\ref{Euler}), (\ref{massEM}) and (\ref{LieMEM}), we can derive a generalization of the Tolman-Oppenheimer-Volkoff (TOV) equations for hydrostatic equilibrium which include anisotropic stresses and electric fields
\begin{align}
\frac{1}{\al}\frac{\da\al}{\da r}=&\frac{M+4\pi\left(\Pmat-2\Pimat\right)R^{3}-\Q^{2}/2R}{R^{2}\left(1-2M/R\right)}\frac{\da R}{\da r}\label{ONE}\\
=&\frac{-1}{\rhomat+\Pmat-2\Pimat}\Bigg[\frac{\da}{\da r}\left(\Pmat-2\Pimat\right)\nonumber\\
&-\frac{1}{R}\frac{\da R}{\da r}\left(6\Pimat-\frac{\Q\ep}{R\sqrt{1-2M/R}}\right)\Bigg],
\end{align}
where the definition for the mass function is now
\begin{align}
\frac{\da M}{\da r}=4\pi\rhomat R^{2}\frac{\da R}{\da r}+\frac{\Q^{2}}{2R^{2}}\frac{\da R}{\da r},
\end{align}
and the energy function is given by
\begin{align}
\left(1+E\right)\left(\frac{\da R}{\da r}\right)^{2}=1-\frac{2M}{R}.\label{frog}
\end{align}
In the static limit, the electric field permeating the spacetime is given by
\begin{align}
\E=\frac{\Q\sqrt{1-2M/R}}{R^{2}}\frac{\da R}{\da r}.
\end{align}
It is trivial to show if we let $R=r$, then from (\ref{static1}) the shift necessarily vanishes, $\be=0$, and all dependance on the temporal coordinate falls out of the equations, reducing to the familiar meaning of a static spacetime.  Furthermore, reducing the energy momentum tensor of the matter to the perfect fluid case ($\Pimat=0$) results in the equations reducing to that of \cite{bekenstein71}.    

\subsection{Exterior static region}
Analogously to the non-static cases, we can utilize the form of the metric for the interior regions of the spacetime to dictate the coordinates for the exterior, vacuum region.  Thus, we again let all the energy-momentum terms vanish at some finite radius, and find the charge and mass respectively become  $\Q=\Q_{0}$ and $M=M_{0}-{\Q_{0}}^{2}/2R$.  Putting this into the equation of hydrostatic equilibrium (\ref{ONE}), we find the lapse function can be integrated to give
\begin{align}
\al=\sqrt{1-\frac{2M_{0}}{R}+\frac{{\Q_{0}}^{2}}{R^{2}}},
\end{align} 
where coordinate freedom has been used to scale an arbitrary function of the temporal coordinate to unity.  There still exists coordinate freedom in the exterior which is determined from a more precise stipulation of the interior.  For example, if we find a solution of the interior region that has $R=r$, this implies through equation (\ref{static1}) that $\be=0$.  Equation (\ref{frog}) then implies $E$ is known, and the metric is simply the RN metric.  Therefore, a static interior charged fluid, with $R=r$, naturally matches onto an RN spacetime in RN coordinates, such that both regions of the spacetime are expressed using the same coordinates.

\section{conclusion}
In this article, we sacrifice mathematical aesthetics for physically interpretable results, and show that many known solutions of the Einstein-Maxwell field equations that are often treated separately can be derived from the same {\it ansatz}.  Furthermore, we showed that the matching of various regions of the spacetime can be done in single coordinate patches as they are all manifestations of the same underlying equations.  We discussed the construction of a qualitative example whereby an initial spacelike hypersurface was established with a general charged fluid interior and vacuum RN spacetime exterior.  This system is evolved using the derived equations and a region of charged Vaidya emerges in a region between the charged fluid and the vacuum.  Therefore, on any $t$ equals constant slice subsequent to the initial hypersurface there exists a region of charged fluid ball extending to some finite radius with charged Vaidya beyond this and vacuum beyond this extending out to spacelike infinity.  This model provides causality to the charged Vaidya solution.

The construction of these solutions using single coordinate patches allowed for the discussion of the collapse of thin shells onto an RN black hole.  A surprising result of this was that a shell of charged matter with the opposite charge to the black hole passes through a region where the ``force'' on the shell was repulsive.  This was shown to be attributed to the geometry of the RN spacetime.   

This work opens avenues for further research into many currently unanswered questions.  In particular the study of the formation and evolution of shell crossing and shell focussing singularities.  While this has been extensively researched for simple fluids models in spherical symmetry, very little is understood about the inclusion of charged fluids.  This work will require numerical analysis as one can quickly see that the equations derived herein are highly coupled, implying closed form solutions are generally not available.  However, the formulation used herein implies this numerical treatment is achievable as the equations are all expressed as an initial value problem.  In particular, in the case of thin charged shells the reduced Einstein-Maxwell differential equations have been reduced to quadrature.    No matching schemes are required at the interfaces between any two regions of the spacetime.  One simply establishes the initial conditions with appropriate equations of states in various regions and the evolution of the system will completely take care of any boundaries.


\bibliography{EM_Paper}

\end{document}